\documentclass[11pt]{article}

\usepackage{sectsty}
\usepackage{graphicx}
\usepackage{listings}
\usepackage{hyperref} 

\title{Comparison of spectrogram scaling in multi-label Music Genre Recognition}
\author{Bartosz Karpiński$^1$, Cyryl Leszczyński$^2$}

\date{\today}

\begin{document}
\maketitle
 
$^1$ Wrocław University of Science and Technology, student at time of writing \\
\texttt{bartosz.karpinski@protonmail.com}

$^2$ Poznań University of Economics and Business, assistant \\\texttt{cyryl.leszczynski@ue.poznan.pl}
\\

\section*{Abstract}
Classifying music into separate genres is an important task, which helps listeners discover new tracks, allows streaming services to better adjust to user preferences, and allows labels and music stores to advertise new albums more effectively. As the accessibility and ease-of-use of digital audio workstations increases, so does the quantity of music available to the public; additionally, differences between genres are not always well defined and can be abstract, with numerous records representing widely varying combinations of genres. In this article, multiple preprocessing methods and approaches to model training are described and compared, accounting for the eclectic nature of today's music. A custom and manually labeled dataset of more than 18000 entries has been used to perform the experiments.

\pagebreak

% Optional TOC
% \tableofcontents
% \pagebreak

%--Paper--

\section{Introduction}
Music genres are a very useful tool to categorize music, which is valuable both for the end user e.g. listener, as they know what other records may be similar to the one they love, as well as for the stores and record labels, as they can try to promote less known records that have similarities to the more popular ones. Thus, the task known as Music Genre Recognition (\textsc{MGR}), in addition to being simply a subset of Audio Classification, has its own important niche of application. 

Tzanetakis and Cook\cite{signal_classification} was one of the first works to show classification of music genres as a pattern recognition task. The authors proposed three sets of characteristics: timbral texture, rhythmic, and pitch content. The classification itself was performed using Statistical Pattern Recognition, Gaussian, Gaussian Mixture Models, and k-Nearest Neighbors. However, the greatest impact was the provided dataset, later named \textsc{GTZAN}, which is a staple of Music Genre Recognition tasks to this day, being often used as a benchmark dataset when performing \textsc{MGR}\cite{ndou2021review}. 

However, \textsc{GTZAN} had many problems, such as simplicity, small number of examples, missing and wrong metadata, and many others\cite{sturm2013gtzan}. This is not to say that the different commonly used data sets did not have a fair share of problems. \textit{One Million Songs Dataset}\cite{bertin2011million}, despite a huge number of examples, only provided tabular data and not raw files, making it unsuitable for use with many of the state-of-the-art approaches. \textit{MagnaTagATune Dataset} lacks proper tagging. Additionally, almost all of them contain only license-free music, generally not as diverse and adequate as popular licensed music, and are far behind the current trends and should be considered outdated and unrepresentative of the current music landscape.

Today, with the advancement of technology, there is an abundance of tools and programs - which are also much easier to use than before - as well as opportunities, and nearly anyone can be an artist. This leads to a massive surge of music released every day. According to the website \textit{Discogs}\cite{discogs}, the number of newly published music increased nearly tenfold from around 400 thousands in the 1950s to more than 4 millions in the 2010s. As the quantity of music available to the public rises, so does the diversity of the records: the genres evolve and blend with others, the differences between them get less well defined and more abstract, with numerous records taking inspiration from several widely varying genres at the same time, or simply mixing them on the record directly. This problem is not addressed in most of the popular datasets, which provide only a single tag per example, greatly simplifying the reality. Thus, a multi-label dataset would be much more beneficial to base the training on, as it is much more fit to handle the multi-style music of today. In this regard, Lukashevich et al. ~\cite{lukashevich2009domain} describe different approaches to multilabeling: for the whole song, by dividing it into separate segments and by dividing into segments while simultaneously dividing into domains, such as Timbre, Rhythm, Melody, and Harmony, with the work being explored and improved further in future papers\cite{nakamura2013borders}. 

Another vital aspect of representing the real-world music landscape would include the inherent imbalance: despite the aforementioned mixes, some genres like Rock, Pop, or Hip-Hop are still overrepresented both in the mainstream and niche compared to others, and those huge disproportions are reflected in the featured or trending lists. Thus, that aspect should also be taken into consideration and adequately reflected in the data used, as with the greater numbers within a genre also comes greater diversity among the examples. Additionally, taking care of adequately handling such imbalance during training should prove useful when finally taking and applying a resulting model in business, as it would face the same - if not even greater - disparities there. In this aspect many of datasets used for research are also seriously lacking - some have a very narrow spectrum of genres, e.g. only classical music, or sometimes containing examples only from a specific country. 

That is why this work is based on the self-aggregated dataset, which consists of music from the 1950s up to the year 2024. Additionally, while it covers a very broad spectrum of genres and styles, there are still many disproportions, resembling the real-world representation and current trends. This allows us to avoid the pitfalls of using other data sets in the research.

The main object of interest of this research is the impact of using different scales of spectrograms during preprocessing on the classification results. Spectrograms are "pictorial representations of the spectrum of frequencies of a signal that varies with time" ~\cite{spectrograms}. Nowadays, they are a basis for most audio processing techniques, utilized not only with Convolutional Neural Networks\cite{pelchat2020neural}\cite{dieleman2011audio}\cite{dhall2021music}, but also with transformers\cite{gong2021ast}\cite{gong2022ssast}\cite{zhu2023multiscale}, and most recently with state space models\cite{lin2024audio}\cite{erol2024audio}\cite{yadav2024audio}\cite{shams2024ssamba}. 

This work focuses on comparing default spectrograms with ones transformed using the Mel scale, brought forward by Stevens et al.\cite{mel_spectrogram} in 1937. This empirically discovered scale defines the range of human-hearable frequencies. It would be only natural that music, intended in purpose for us humans to hear, would benefit from representing it using the Mel scale. 

\section{Related works}
Bahuleyan\cite{bahuleyan2018ml} presents the usage of CNN models in spectrograms, machine learning models in time domain features, and the combination of both in the task of music genre classification. Here also the VGG16 was used, and for the Machine Learning Algorithms the Linear Regression, Gradient Boosting, Random Forest, and Support Vector Machines. The combination of VGG16 and Gradient Boosting was also used as an ensemble, which proved successful as it achieved the best result of all the other compared methods. 

In Ali and Siddiqui\cite{ali2017automatic} the authors used k-NN and SVM techniques on the Mel Frequency Cepstral Coefficients extracted again from the GTZAN dataset, with and without dimensionality reduction. Dhall et al.\cite{dhall2021music} compares the F-transform, Q-transform, and mel spectrograms also on GTZAN dataset using Convolutional Neural Networks, showing small differences between the methods when used with grayscale, and mel spectrograms perform significantly better for RGB images. The same data set was used by KM et al.\cite{km2021deep}, where the authors proposed a CNN-based classification that would be an input to a recommendation system. Also using the GTZAN dataset, Dong\cite{dong2018convolutional} proposes a model that divides mel spectrograms into 3-second segments before being put into a CNN, reasoning that the accuracy of human classification plateaus at that mark. Pelchat and Gelowitz\cite{pelchat2020neural} use a similar approach, feeding a neural network with small spectrogram slices of less than 3 seconds. Mehta et al. ~\cite{mehta2021music} use transfer learning using Resnet, but also VGG and AlexNet, on the GTZAN dataset using log-based mel spectrograms. Matocha and Zieliński\cite{matocha2018music} compare the usage of monoaural signals and stereophonic signals as input to CNNs, where the latter performed worse, to the surprise of the authors. Yang and Zhang\cite{yang2019music} utilized a duplicated convolutional architecture to split the input for different pooling layers, to gather more statistical information used in the classification process. 

Nirmal and Mohan\cite{nirmal2020spectrograms} compared a self-made CNN architecture with a pre-trained MobileNet model. Costa et al.\cite{costa2017spectrogram} test representation learning in tandem with Convolutional Networks. In addition to using the spectrograms to perform the classification task, the authors also experimented with the fusion of both learned and handcrafted features to assess the complementarity of those two approaches. Zhao et al.\cite{zhao2022s3t} proposed a novel Swin Transformer architecture for music classification, using a self-supervised approach and augmentation of spectrograms, by applying multi-crop, frequency or time masking, shifting or warping at random. Choi et al.\cite{choi2017crnn} shows some strong performances utilizing Convolutional Recurrent Neural Networks, where the last convolutional layers are replaced with the Recurrent Neural Network model. Schindler and Knees\cite{schindler2019multi} tackle the multi-label problem of tagging music by features such as genre, style, mood or theme. In the experiments they've used the the Million Song Dataset, and their Deep Neural Models were based on the triplet-network architecture. The same dataset was the basis for Dieleman et al.\cite{dieleman2011audio}, where, due to the sheer size of the dataset, the authors had to resort to training a deep convolutional belief network on the entire dataset and then train a multilayer perceptron with the parameters learned from the previous network for initialization. 

Silla et al.\cite{silla2008automl} proposes the combination of using the time and space decomposition with ensemble classifiers, using classical machine learning methods such as K-Nearest Neighbors, Naïve-Bayes, Decision Trees and Support Vector Machines on a novel Latin dataset. In Oramas et al.\cite{oramas2017multilabel}, the authors present a comparison of different feature extraction methodologies from audio, text reviews, and image covers, as well as their combinations, showing ways to improve accuracy by implementing approaches that may not sound so obvious at first. Here, the CNN is used on the spectrograms for audio, a Vector Space Model for the text, and a modified version of the ResNet model for the covers. The results show the potential of the multimodal approach. Sanden and Zhang\cite{sanden2011enhancing} discuss different ensemble techniques for the specific task of the multi-label classification of music genres. Although using only a simpler machine learning approach, the whole research is very robust, covering multiple datasets and ensemble techniques. 

\section{Experimental evaluation}
This section describes in detail the experimental evaluation process conducted in order to analyze the impact of different spectrogram preprocessing on the performance of different Convolutional Neural Networks with the use of transfer learning. The experiment performed was prepared to answer the following questions:

– Does the use of Mel-scaled spectrograms offer a statistically better performance using CNNs with Transfer Learning in comparison to unscaled spectrograms?

– Does the difference (or lack thereof) stay the same between different ResNet implementations that vary in depth?

\subsection{Set-up}

\paragraph{Dataset}The data used for the experiment are derived from the manually aggregated, private dataset of songs in MP3, MP4 or FLAC format. The relevant tags for the songs consist of up to three references to a subgenre tag, which is a fine-grain defined genre, which in turn has reference to up to 2 broader-defined genre. To provide an example, a subgenre would be electro-industrial, which is quite a niche genre in itself (only 50 records belong to it). Furthermore, it itself belongs to the categories "electronic" and "industrial \& noise". The tags were scraped from different on-line music databases, which are maintained by large music communities.

\begin{table}[!htbp]
\centering\caption{Number of songs by genre \label{table:genres}}
\begin{tabular}{|l|l|l|l|}% table alignment -> l c r - left, center, right
\hline
Hip-Hop	& 5641  & R\&B	& 801 \\
\hline
Electronic & 4680 & Industrial \& Noise	& 788 \\
\hline
Rock & 4356 & Experimental & 694 \\
\hline
Pop	& 2758 & Ambient & 687 \\
\hline
Psychedelia	& 1367 & Folk & 652 \\
\hline
Metal & 1153 & Classical Music	& 608 \\
\hline
Dance & 960 & Singer-Songwriter & 474 \\
\hline
Punk & 804 & Jazz & 278 \\
\hline
\end{tabular}
\end{table}

In table \ref{table:genres} all records from the Genres table are listed, together with the number of songs that are associated with each one. A huge disproportion between some of the genres can be seen, as the largest one contains 5641 records associated with it, whilst the smallest one only 23. Due to this huge disproportion, only 16 Genres were chosen: from hip-hop down to jazz. Whilst Regional Music is almost similar size as jazz, it is a very broad genre encompassing a very diverse range of music from all over the world. Its classification is a difficult task in itself ~\cite{lukashevich2009domain}. In total, 18019 data samples were provided, because some of them are represented in a few genres.

\paragraph{Data preprocessing}

\begin{figure}[!htbp]
    \centering
    \begin{minipage}{0.35\textwidth}
        \centering
        \includegraphics[width=\linewidth]{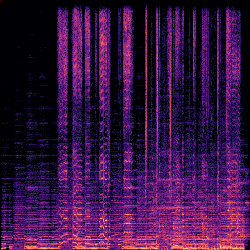}
        \caption{A sample default spectrogram}
        \label{fig:default_normal}
    \end{minipage}\hfill
    \begin{minipage}{0.35\textwidth}
        \centering
        \includegraphics[width=\linewidth]{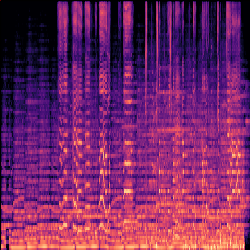}
        \caption{A sample mel-spectrogram of the same song}
        \label{fig:default_mel}
    \end{minipage}
\end{figure}

The data for the experiment come in two forms: spectrograms (Figure \ref{fig:default_normal}), as well as mel scale spectrograms (mel spectrograms) (Figure \ref{fig:default_mel}). Both were obtained using the Librosa Python library. The first is a visual representation of signal frequencies over time, while the latter is a spectrogram specially adjusted to the Mel scale, which is a perceptual scale adjusted to human hearing ~\cite{mel_spectrogram}.

\paragraph{Model architecture}

The basis of the classification was an ensemble of binary classifiers, each for one genre, trained using a \textit{One-vs-All (OVA)} binarization strategy. Through this ensemble a multilabel genre classificator was achieved. The base models used were different types of ResNet, namely \textit{ResNet34}, \textit{Resnet50}, \textit{ResNet101}, and lastly \textit{ResNet152}. During training, the transfer learning approach was chosen, for which \textit{ImageNetV1} weights were used as a basis.

\subsection{Experimental scenario}

\textbf{Scenario} Comparison of normal and mel spectrograms with a stratified subset derived from the negative class, equal in size to the positive class, for each model and so for each genre individually during training. This subset was then stationary, i.e. it was not changed between epochs as each and every one of them used it as a whole. For testing, a stratified subset equal to 10\% of the entire data set was used.
\textbf{Goals} Comparison of the performance of both spectrograms and mel spectrograms across \textit{Resnet} models with various layer depths.

\subsection{Evaluation Protocol}
Due to the size of the data set and the number of models that had to be trained, it was ultimately decided not to perform any cross-validation. On the basis of one run, several metrics were used. For singular-genre models: Accuracy, Loss, F1 Score, Balanced Accuracy, Precision, and Recall.
% , Specificity, Geometric Mean, and Compute Area Under the Receiver Operating Characteristic Curve. For the multi-label ensemble, all of the previously mentioned were used, and additionally Hamming Loss and Strict Ensemble Accuracy (On the graphs shown as "0/1"), being a binary evaluation of the ensemble - if all models were correct or not. 

\subsection{Reproducibility}
 The entirety of the code was written in \textit{Python} 3.12.7. with CUDA version 12.4. In terms of libraries, the main ones used were \textit{Pytorch} 2.5.1\cite{paszke2019pytorch}, \textit{Librosa} 0.10.2\cite{mcfee2015librosa}, \textit{scikit-learn} 1.5.2\cite{pedregosa2011scikit}, \textit{Pandas} 2.2.3\cite{mckinney-proc-scipy-2010}, and \textit{Seaborn} 0.13.2\cite{Waskom2021}. The computing platform included 64GB of RAM, a 16-core CPU, and 8448 CUDA cores GPU with 16GB of VRAM. The entire code is available on a \textit{Github} repository\footnote{\href{https://github.com/CaptainBoii/spectrograms-comparison}{https://github.com/CaptainBoii/spectrograms-comparison}}

\section{Results}

\begin{figure}[!htbp]
    \centering
    \includegraphics[width=0.8\linewidth]{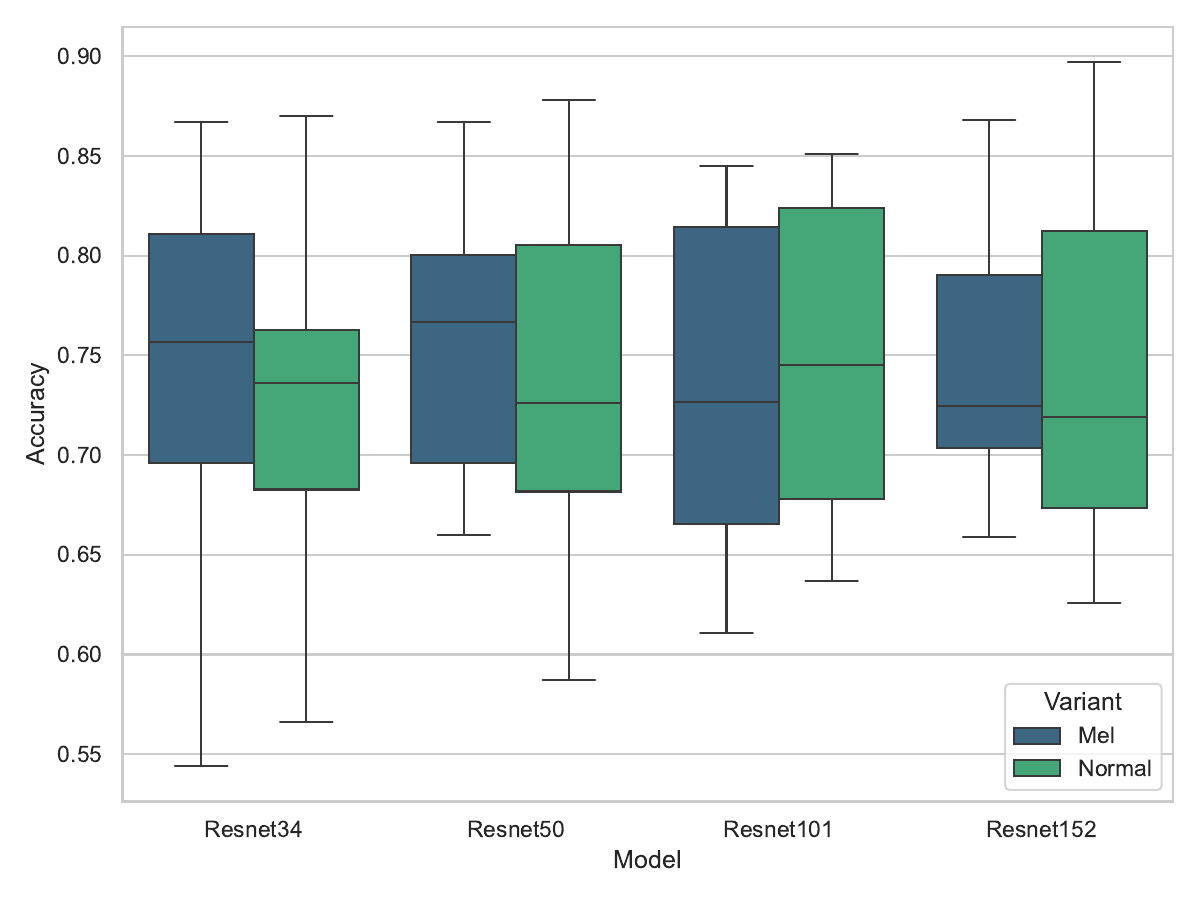}
    \caption{Average accuracy across models}
    \label{fig:avg_accuracy}
\end{figure}

\begin{figure}[ht!]
    \centering
    \includegraphics[width=0.8\linewidth]{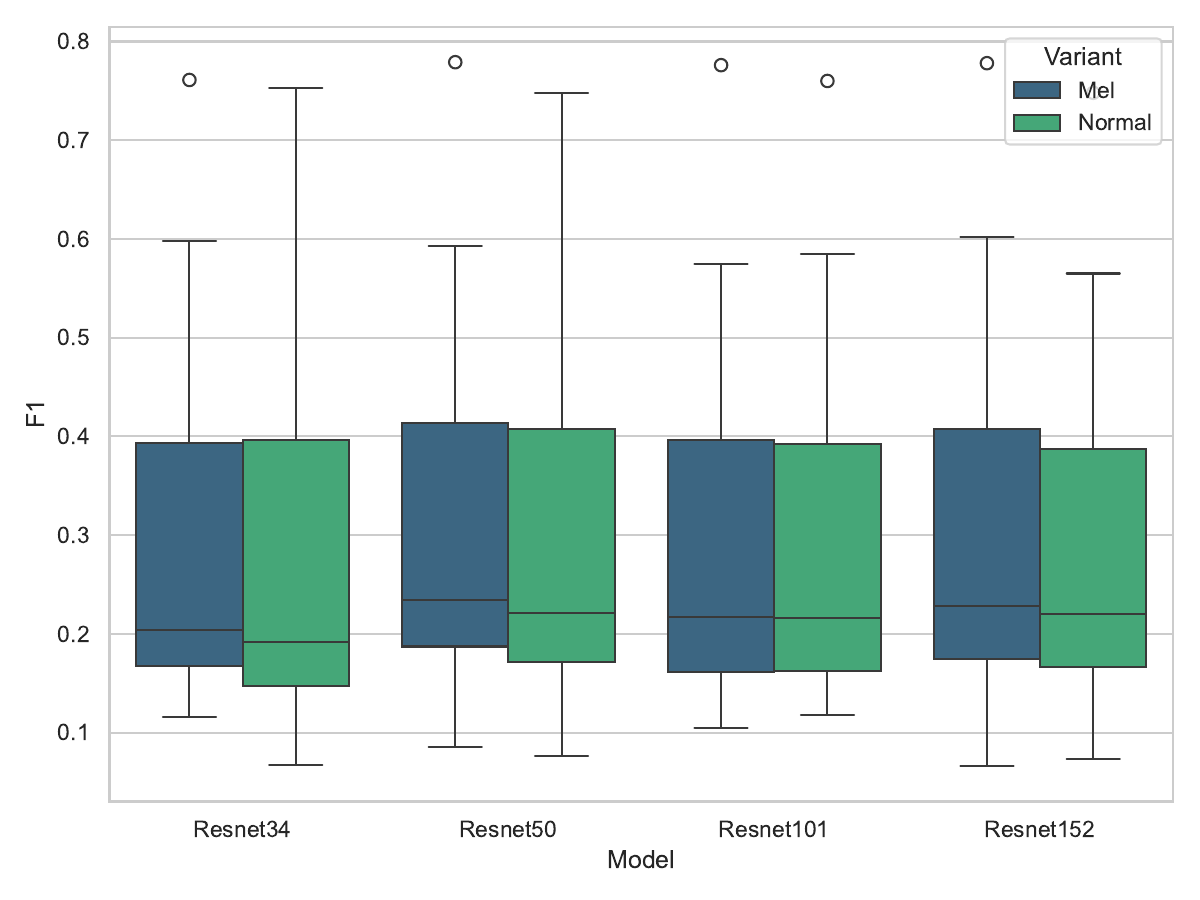}
    \caption{Average F1 Score across models}
    \label{fig:avg_f1}
\end{figure}

\begin{figure}[!htbp]
    \centering
    \includegraphics[width=0.8\linewidth]{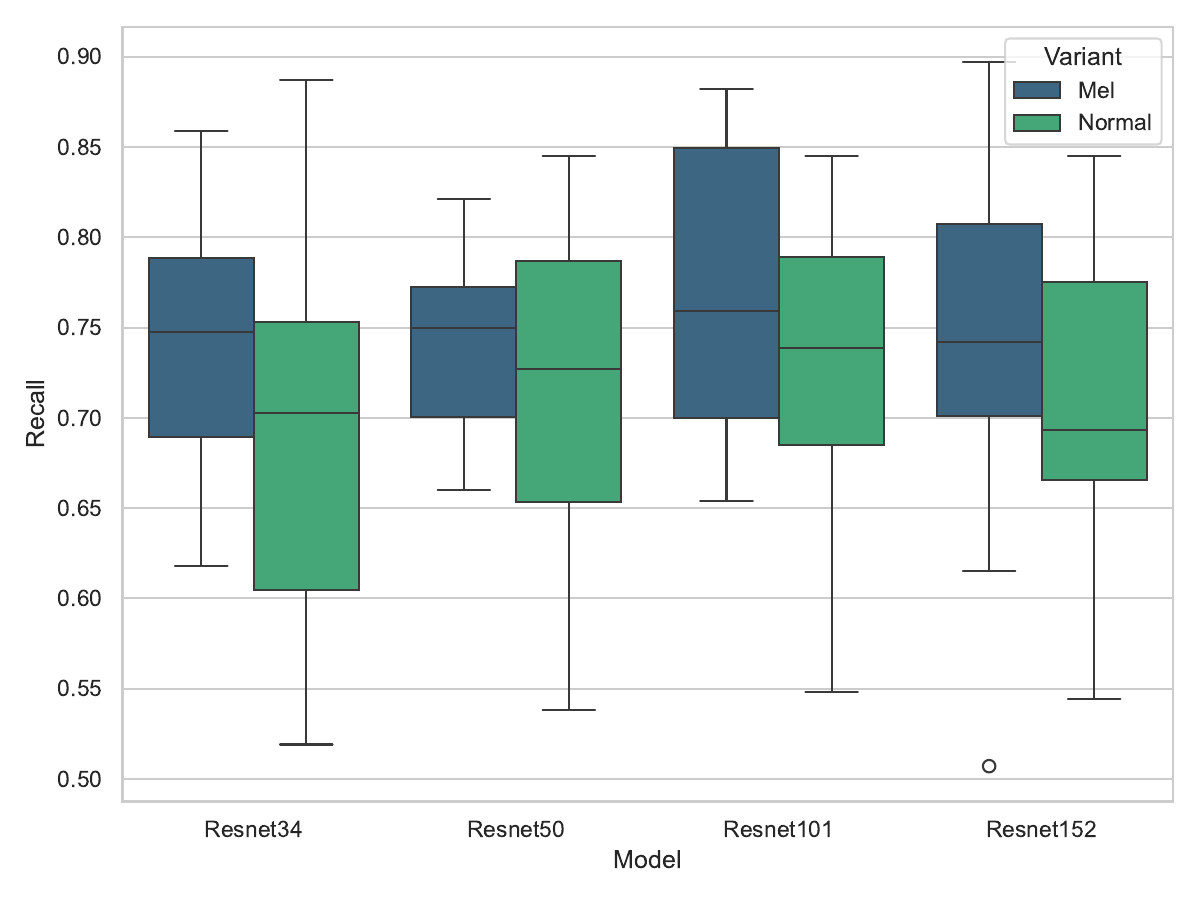}
    \caption{Average Recall across models}
    \label{fig:avg_recall}
\end{figure}

\begin{figure}[!htbp]
    \centering
    \includegraphics[width=0.8\linewidth]{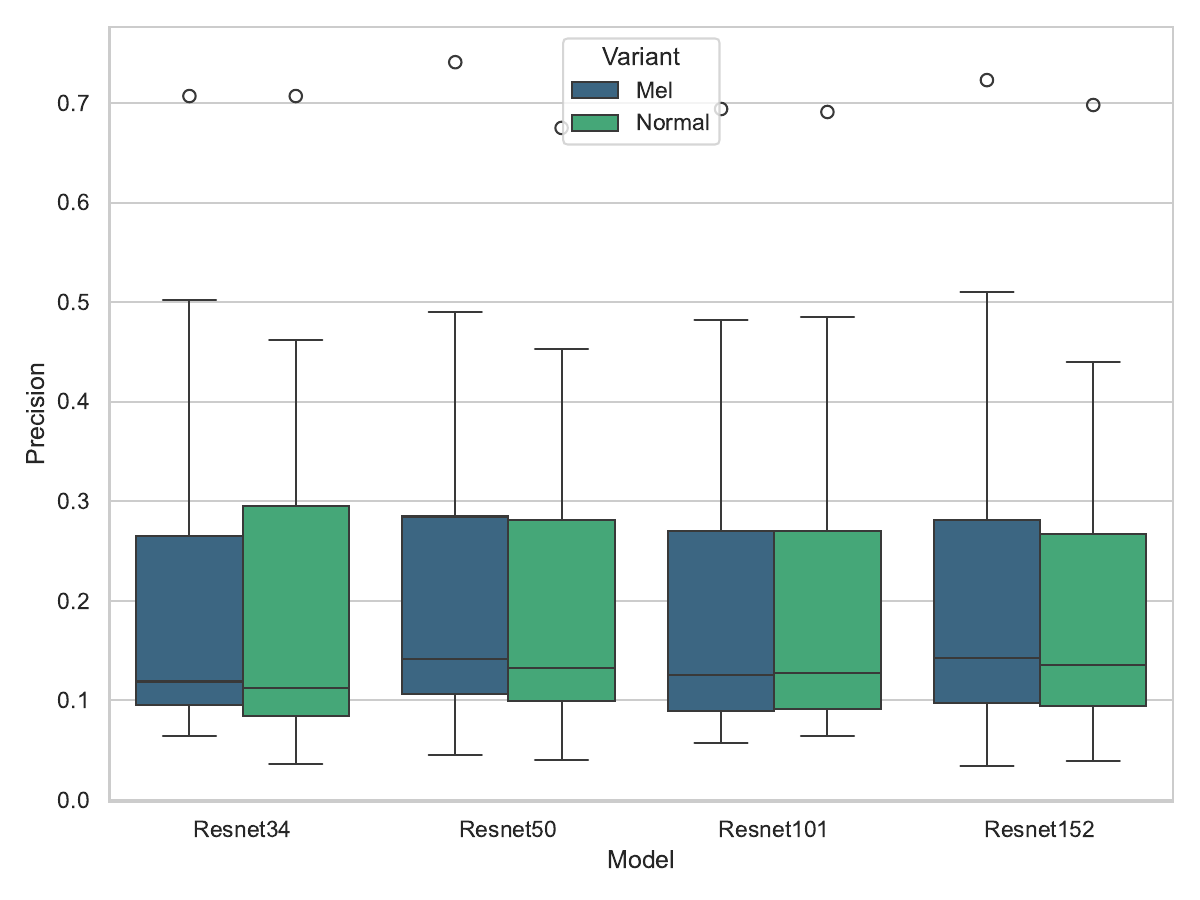}
    \caption{Average precision across models}
    \label{fig:avg_precision}
\end{figure}

Starting with models trained without transfer learning, Figure \ref{fig:avg_accuracy} shows a rather distinctive difference between normal and scaled spectrograms. In this case, only \textit{Resnet101} appears to have a better median. Nothing conclusive could be said about variance as neither of the approaches seems to be better across all model depths. More interesting is Figure \ref{fig:avg_f1}, where the median is universally better when using the mel spectrogram approach. The variance is also noticeably greater in the standard approach, although it results in having more having higher-scoring models. Despite notably lower than raw accuracy, here the scores tell us about the problem in a clearer way, as they more take into account the inherent imbalance of the dataset. In Figure \ref{fig:avg_recall}, it tells a similar story. The mel spectrograms perform better, with both higher medians and distribution skewed more towards the top, with the bottom quantiles also being much more top-leaning. The worst performance can be observed in Figure \ref{fig:avg_precision}, with neither of the approaches performing better than the other. This can be attributed to the stratified nature of the test subset, as some genres had up to 20 times more samples than the others, which makes it not ideal for models to be truly precise. 



\begin{figure}[!htbp]
  \centering
  \includegraphics[width=1\linewidth]{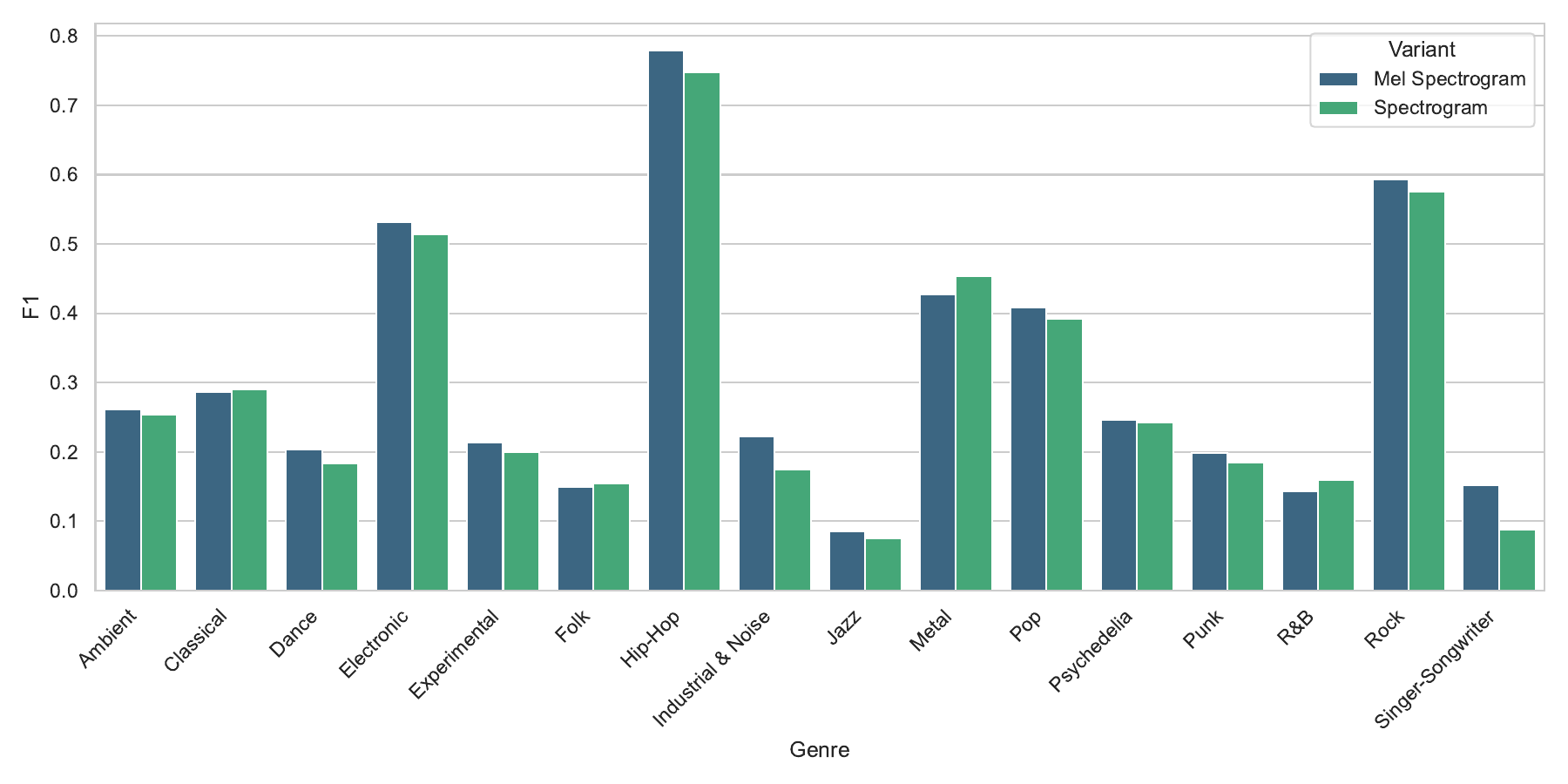}
  \caption{Average F1 Score across genres for Resnet50}
  \label{fig:genres_f1}
\end{figure}

\begin{figure}[!htbp]
  \centering
  \includegraphics[width=1\linewidth]{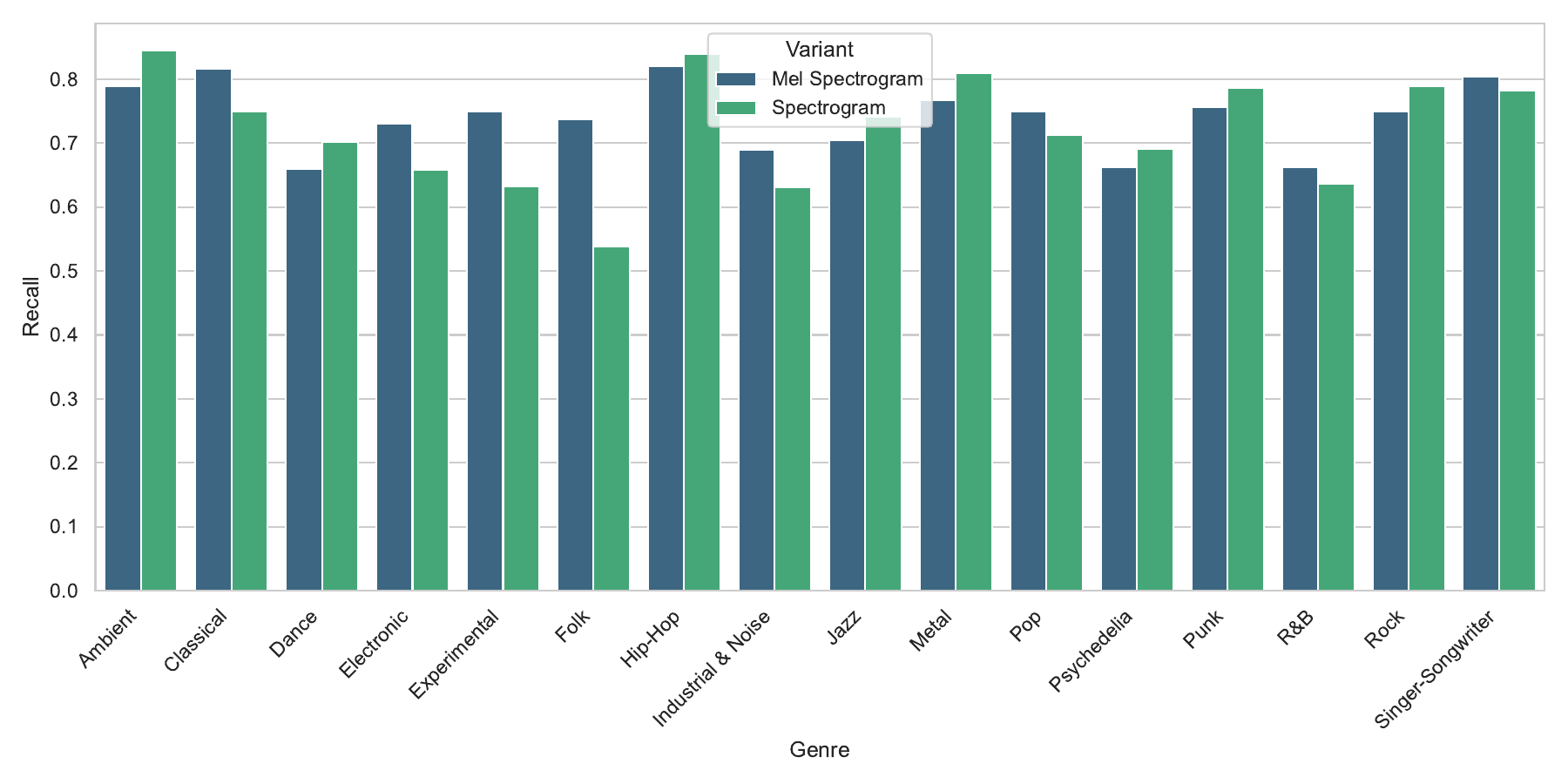}
  \caption{Average Recall across genres for Resnet50}
  \label{fig:genres_recall}
\end{figure}

\begin{figure}[!htbp]
  \centering
  \includegraphics[width=1\linewidth]{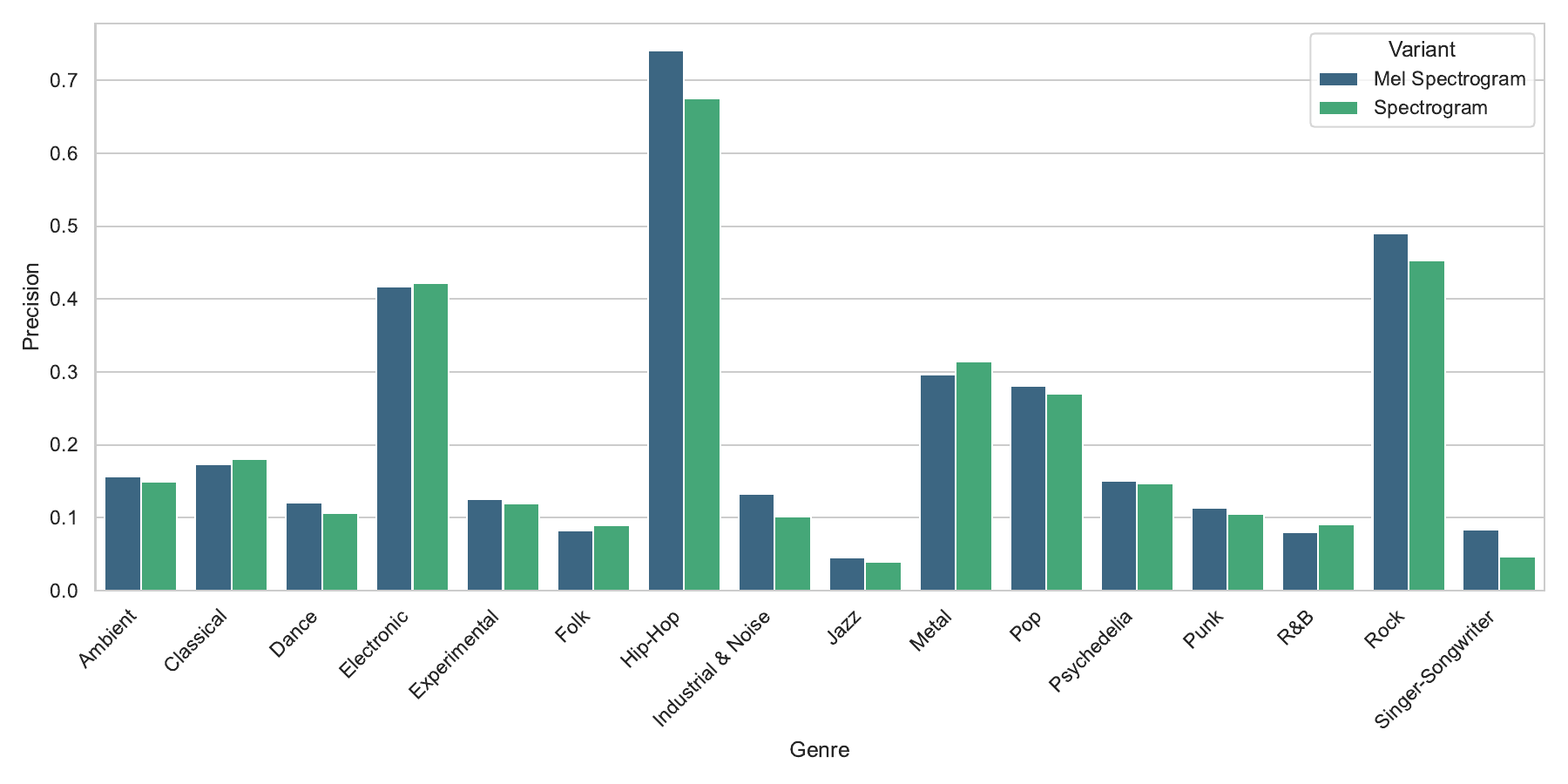}
  \caption{Average Precision across genres for Resnet50}
  \label{fig:genres_precision}
\end{figure}

In Figure \ref{fig:genres_f1} we can see huge disproportions between the genres. Interestingly, higher scores appear to strongly correlate with the sizes of the genres \ref{table:genres}. As Recall seems to be fairly uniform across all cases in Figure \ref{fig:genres_recall}, taking a look at Figure \ref{fig:genres_precision} gives us more information. Again, the high correlation with the sizes of the subsets seems to explain the scores of the F1, which is a harmonic mean of Precision and Recall, and as the latter does not change much across the genres, the resulting F1 is heavily dependent on the Precision. Thus, this correlation can be easily explained with the size distributions, as the greater the size of the subset, the better a model could be trained on it. In return, this means that providing more examples for the underrepresented classes would probably yield the best improvements to the overall score, as it would greatly boost both average and median. As for the differences between mel spectrograms and normal ones, the biggest difference seems to be between two of the largest subsets: Hip Hop and Rock. This might imply that increasing the size of the other genres' subset would, in fact, improve the difference between the two types.


In order to assess the impact of using mel-scale and non-mel-scale spectrograms during the experiments, the F1 metric has been used as a point of comparison between the two. Aggregated performance values from transfer models have been considered in order to evaluate whether the choice of spectrogram type yields statistically different results.

For each type of spectrogram, aggregated F1 scores were obtained across the models. Because the aggregated values represent paired observations (that is, each model provides an F1 score for both spectrogram types), the difference in F1 scores (spectrogram F1 minus mel spectrogram F1) has been calculated for each model.

\begin{figure}[!htbp]
    \centering
    \includegraphics[width=0.75\linewidth]{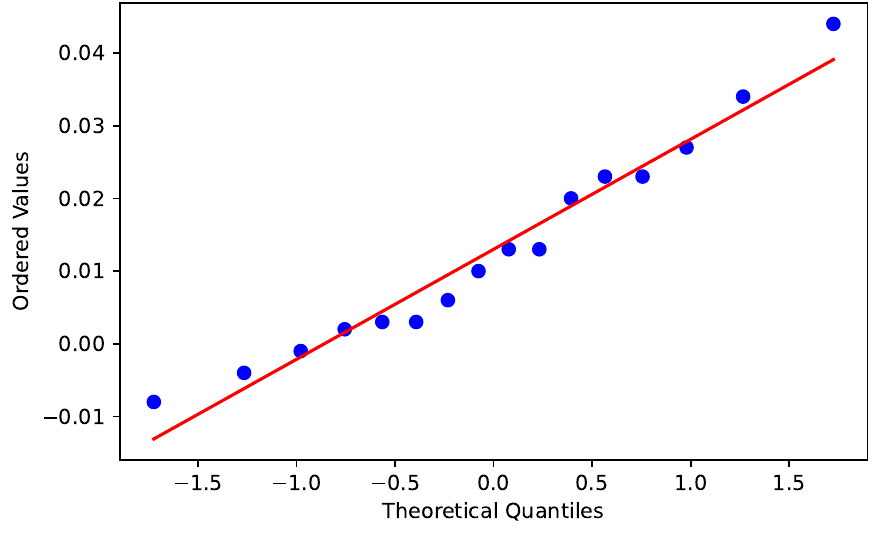}
    \caption{Q–Q Plot for Differences in F1 between mel spectrograms and normal spectrograms}
    \label{fig:qq}
\end{figure}

Before performing the main hypothesis test, the normality of these differences was assessed using the Shapiro–Wilk test, which failed to reject the null hypothesis of normality - as such, the differences appear approximately normally distributed. The Q-Q plot in Figure \ref{fig:qq} shows the differences against the theoretical quantiles of a "perfectly" normal distribution, showing only modest variations towards the extremes (which indicates normality). This confirmation of normality, combined with the paired nature of the compared data, narrowed the selection of the test to a paired t-test.

% To jest tłumaczenie się z przyjęcia p na poziomie 0.1 zamiast 0.05
% Given that the experimental results are derived from a limited number of aggregate observations, the analysis is susceptible to Type II errors. Due to the low number of observations, a conventional $\alpha=0.05$ increases the risk of overlooking a meaningful difference. Due to the exploratory nature of the study and in an attempt to balance the risk of error against Type II errors, the decision to set $\alpha$ at $0.1$ has been made.

% Under this assumption, the results of the t-test ($p=0.08$) indicate that there is a statistically significant difference between the two types of spectrograms in terms of their F1 scores for this type of highly varied dataset. The resulting t-statistic of $-1.863$, indicates that mel spectrograms have a slight edge over normal scale spectrograms in terms of F1 scores, which matches the initial expectations.

% Po aktualizacji wyników - mamy dla porównania malutkie p
The results of the dependent samples t test ($p=0.0027$) indicate that there is a statistically significant difference between the two types of spectrograms, as far as their F1 scores are concerned in this highly varied dataset. The resulting t-statistic of $-3.5872$ is indicative of mel spectrograms that have an edge over normal scale spectrograms (in terms of F1 scores), which matches the a priori expectations of the experiment results.

% \section{Analysis}
% % Realnie to nie będzie sekcja, a raczej kawałek w conclusions albo results czymś takim. Redakcyjnie do wypracowania. Zamieszczam w osobnym pliku, żeby łatwiej było wyjmować relewantne informacje do finalnego artykułu.
% \input{analysis}

\section{Conclusions and future research}
Both through the results and the statistical analysis, it was shown that the mel spectrograms perform better in the music classification task, so often used in the field of the Music Information Retrieval. It shows how the preprocessing of the audio files leaves a lot of space for future improvements, and comparison with other scales of spectrograms (like Q and F) and tuning with the spectrograms parameters may further yield better results. Additionally, different libraries (like \textit{Librosa}, \textsc{FFMPEG}) seem to have different back-end implementations, so despite being the same spectrograms in theory, the images might look vastly different. Also, some adjustments may also prove useful when trying to tailor the data specifically for the state-of-the-art architectures like the Audio Spectrogram Transformer\cite{gong2021ast} and its derivatives\cite{gong2022ssast}\cite{zhu2023multiscale} or Mamba-based ones\cite{lin2024audio}\cite{erol2024audio}\cite{yadav2024audio}\cite{shams2024ssamba}, which for now simply use regular spectrograms under the hood. Another vital aspect of this work is bringing forward a new dataset that through the achieved scores also leaves a lot of room for improvement in the future, while also bringing more detailed labels, more samples, and enables a multi-label approach. The analysis of F1 Scores and Precision across genres indicates the strong correlation between genre subset size and it's precision, and in result the F1 Score, so adjusting the ratio and collecting more data will probably increase both the accuracy and the need for a special approach towards imbalance. Hopefully, after further data gathering and other improvements, this dataset may prove an interesting challenge for other researchers to tackle, and even become a new benchmark dataset for Music Genre Classification, replacing the overused and unsuitable GTZAN.

%--/Paper--
\newpage

% -- Refs --
\bibliographystyle{bibliografia}
\bibliography{bibliography}

\begin{thebibliography}{10}

\bibitem{ali2017automatic}
Ali, M.A., Siddiqui, Z.A., \emph{Automatic music genres classification using machine learning}, International Journal of Advanced Computer Science and Applications. 2017, vol.~8, 8.

\bibitem{bahuleyan2018ml}
Bahuleyan, H., \emph{Music genre classification using machine learning techniques}, arXiv preprint arXiv:1804.01149. 2018.

\bibitem{bertin2011million}
Bertin-Mahieux, T., Ellis, D.P., Whitman, B., Lamere, P., \emph{The million song dataset.}, in: \emph{Ismir}, 9 (2011), page~10.

\bibitem{choi2017crnn}
Choi, K., Fazekas, G., Sandler, M., Cho, K., \emph{Convolutional recurrent neural networks for music classification}, in: \emph{2017 IEEE International conference on acoustics, speech and signal processing (ICASSP)} (IEEE, 2017), pp. 2392--2396.

\bibitem{costa2017spectrogram}
Costa, Y.M., Oliveira, L.S., Silla~Jr, C.N., \emph{An evaluation of convolutional neural networks for music classification using spectrograms}, Applied soft computing. 2017, vol.~52, pp. 28--38.

\bibitem{dhall2021music}
Dhall, A., Srinivasa~Murthy, Y., Koolagudi, S.G., \emph{Music genre classification with convolutional neural networks and comparison with f, q, and mel spectrogram-based images}, in: \emph{Advances in Speech and Music Technology: Proceedings of FRSM 2020} (Springer, 2021), pp. 235--248.

\bibitem{dieleman2011audio}
Dieleman, S., Brakel, P., Schrauwen, B., \emph{Audio-based music classification with a pretrained convolutional network}, in: \emph{12th International Society for Music Information Retrieval Conference (ISMIR-2011)} (University of Miami, 2011), pp. 669--674.

\bibitem{dong2018convolutional}
Dong, M., \emph{Convolutional neural network achieves human-level accuracy in music genre classification}, arXiv preprint arXiv:1802.09697. 2018.

\bibitem{erol2024audio}
Erol, M.H., Senocak, A., Feng, J., Chung, J.S., \emph{Audio mamba: Bidirectional state space model for audio representation learning}, IEEE Signal Processing Letters. 2024.

\bibitem{spectrograms}
French, M., Handy, R., \emph{Spectrograms: turning signals into pictures}, Journal of Engineering Technology, vol. 24, pp. 32-35. 2007.

\bibitem{gong2021ast}
Gong, Y., Chung, Y.A., Glass, J., \emph{Ast: Audio spectrogram transformer}, arXiv preprint arXiv:2104.01778. 2021.

\bibitem{gong2022ssast}
Gong, Y., Lai, C.I., Chung, Y.A., Glass, J., \emph{Ssast: Self-supervised audio spectrogram transformer}, in: \emph{Proceedings of the AAAI Conference on Artificial Intelligence}, vol.~36 (2022), pp. 10699--10709.

\bibitem{km2021deep}
KM, A., i~in., \emph{Deep learning based music genre classification using spectrogram}, in: \emph{Proceedings of the International Conference on IoT Based Control Networks \& Intelligent Systems-ICICNIS} (2021).

\bibitem{lin2024audio}
Lin, J., Hu, H., \emph{Audio mamba: Pretrained audio state space model for audio tagging}, arXiv preprint arXiv:2405.13636. 2024.

\bibitem{lukashevich2009domain}
Lukashevich, H.M., Abe{\ss}er, J., Dittmar, C., Grossmann, H., \emph{From multi-labeling to multi-domain-labeling: A novel two-dimensional approach to music genre classification.}, in: \emph{ISMIR} (2009), pp. 459--464.

\bibitem{matocha2018music}
Matocha, M., Zieli{\'n}ski, S., \emph{Music genre recognition using convolutional neural networks}, Advances in Computer Science Research. 2018.

\bibitem{mcfee2015librosa}
McFee, B., Raffel, C., Liang, D., Ellis, D.P., McVicar, M., Battenberg, E., Nieto, O., \emph{librosa: Audio and music signal analysis in python.}, SciPy. 2015, vol. 2015, pp. 18--24.

\bibitem{mckinney-proc-scipy-2010}
{W}es {M}c{K}inney, \emph{{D}ata {S}tructures for {S}tatistical {C}omputing in {P}ython}, in: \emph{{P}roceedings of the 9th {P}ython in {S}cience {C}onference}, pod red. {S}t\'efan van~der {W}alt, {J}arrod {M}illman (2010), pp. 56 -- 61.

\bibitem{mehta2021music}
Mehta, J., Gandhi, D., Thakur, G., Kanani, P., \emph{Music genre classification using transfer learning on log-based mel spectrogram}, in: \emph{2021 5th International Conference on Computing Methodologies and Communication (ICCMC)} (IEEE, 2021), pp. 1101--1107.

\bibitem{nakamura2013borders}
Nakamura, H., Huang, H.H., Kawagoe, K., \emph{Detecting musical genre borders for multi-label genre classification}, in: \emph{2013 IEEE International Symposium on Multimedia} (IEEE, 2013), pp. 532--533.

\bibitem{ndou2021review}
Ndou, N., Ajoodha, R., Jadhav, A., \emph{Music genre classification: A review of deep-learning and traditional machine-learning approaches}, in: \emph{2021 IEEE International IOT, Electronics and Mechatronics Conference (IEMTRONICS)} (IEEE, 2021), pp. 1--6.

\bibitem{nirmal2020spectrograms}
Nirmal, M., Mohan, S., \emph{Music genre classification using spectrograms}, in: \emph{2020 International conference on power, instrumentation, control and computing (PICC)} (IEEE, 2020), pp. 1--5.

\bibitem{oramas2017multilabel}
Oramas, S., Nieto, O., Barbieri, F., Serra, X., \emph{Multi-label music genre classification from audio}, Text, and Images Using Deep Features. 2017, vol.~21.

\bibitem{paszke2019pytorch}
Paszke, A., \emph{Pytorch: An imperative style, high-performance deep learning library}, arXiv preprint arXiv:1912.01703. 2019.

\bibitem{pedregosa2011scikit}
Pedregosa, F., Varoquaux, G., Gramfort, A., Michel, V., Thirion, B., Grisel, O., Blondel, M., Prettenhofer, P., Weiss, R., Dubourg, V., i~in., \emph{Scikit-learn: Machine learning in python}, the Journal of machine Learning research. 2011, vol.~12, pp. 2825--2830.

\bibitem{pelchat2020neural}
Pelchat, N., Gelowitz, C.M., \emph{Neural network music genre classification}, Canadian Journal of Electrical and Computer Engineering. 2020, vol.~43, 3, pp. 170--173.

\bibitem{sanden2011enhancing}
Sanden, C., Zhang, J.Z., \emph{Enhancing multi-label music genre classification through ensemble techniques}, in: \emph{Proceedings of the 34th international ACM SIGIR conference on Research and development in Information Retrieval} (2011), pp. 705--714.

\bibitem{schindler2019multi}
Schindler, A., Knees, P., \emph{Multi-task music representation learning from multi-label embeddings}, in: \emph{2019 International Conference on Content-Based Multimedia Indexing (CBMI)} (IEEE, 2019), pp. 1--6.

\bibitem{shams2024ssamba}
Shams, S., Dindar, S.S., Jiang, X., Mesgarani, N., \emph{Ssamba: Self-supervised audio representation learning with mamba state space model}, in: \emph{2024 IEEE Spoken Language Technology Workshop (SLT)} (IEEE, 2024), pp. 1053--1059.

\bibitem{silla2008automl}
Silla, C.N., Koerich, A.L., Kaestner, C.A., \emph{A machine learning approach to automatic music genre classification}, Journal of the Brazilian Computer Society. 2008, vol.~14, pp. 7--18.

\bibitem{mel_spectrogram}
Stevens, S.S., Volkmann, J., Newman, E.B., \emph{A scale for the measurement of the psychological magnitude pitch}, The journal of the acoustical society of america. 1937, vol.~8, 3, pp. 185--190.

\bibitem{sturm2013gtzan}
Sturm, B.L., \emph{The gtzan dataset: Its contents, its faults, their effects on evaluation, and its future use}, arXiv preprint arXiv:1306.1461. 2013.

\bibitem{signal_classification}
Tzanetakis, G., Cook, P., \emph{Musical genre classification of audio signals}, IEEE Transactions on speech and audio processing. 2002, vol.~10, 5, pp. 293--302.

\bibitem{Waskom2021}
Waskom, M.L., \emph{seaborn: statistical data visualization}, Journal of Open Source Software. 2021, vol.~6, 60, page 3021.

\bibitem{yadav2024audio}
Yadav, S., Tan, Z.H., \emph{Audio mamba: Selective state spaces for self-supervised audio representations}, arXiv preprint arXiv:2406.02178. 2024.

\bibitem{yang2019music}
Yang, H., Zhang, W.Q., \emph{Music genre classification using duplicated convolutional layers in neural networks.}, in: \emph{Interspeech} (2019), pp. 3382--3386.

\bibitem{zhao2022s3t}
Zhao, H., Zhang, C., Zhu, B., Ma, Z., Zhang, K., \emph{S3t: Self-supervised pre-training with swin transformer for music classification}, in: \emph{ICASSP 2022-2022 IEEE International Conference on Acoustics, Speech and Signal Processing (ICASSP)} (IEEE, 2022), pp. 606--610.

\bibitem{zhu2023multiscale}
Zhu, W., Omar, M., \emph{Multiscale audio spectrogram transformer for efficient audio classification}, in: \emph{ICASSP 2023-2023 IEEE International Conference on Acoustics, Speech and Signal Processing (ICASSP)} (IEEE, 2023), pp. 1--5.

\bibitem{discogs}
{Zink Media Inc.}, \emph{Discogs search page}, \url{https://www.discogs.com/search/}.

\end{thebibliography}

\end{document}